\documentclass[10pt,twocolumn]{article}

\usepackage[utf8]{inputenc}
\usepackage[T1]{fontenc}
\usepackage{lmodern}
\usepackage{amsmath,amssymb}
\usepackage{graphicx}
\usepackage[letterpaper,margin=0.7in]{geometry}
\usepackage[super,sort&compress]{natbib}
\usepackage{authblk}
\usepackage[font=small]{caption}
\usepackage{microtype}
\usepackage{xcolor}
\usepackage[bookmarks=false]{hyperref}

\ifdefined\pdfsuppresswarningpagegroup \pdfsuppresswarningpagegroup=1 \fi

\renewcommand{\eqref}[1]{\textup{Eq.~\ref{#1}}}

\makeatletter
\let\SIsection\section
\renewcommand\section{\@startsection{section}{1}{\z@}%
  {-3.5ex \@plus -1ex \@minus -.2ex}{2.3ex \@plus.2ex}{\normalfont\Large\bfseries\raggedright}}
\renewcommand\paragraph{\@startsection{paragraph}{4}{\parindent}%
  {3.25ex \@plus1ex \@minus.2ex}{-1em}{\normalfont\normalsize\bfseries\boldmath}}
\makeatother

\hypersetup{
  pdftitle={Soft Matter, Hard Rules: Emulsions Follow the Laws of Granular Suspension Rheology},
  pdfauthor={Wenjun Chen, Eric De Giuli, Matthieu Wyart, Yo\"el Forterre, Jasna Bruji\'c, Bloen Metzger}
}

\title{\textbf{Soft Matter, Hard Rules:\\ Emulsions Follow the Laws of Granular Suspension Rheology}}

\author[1,2]{Wenjun Chen}
\author[3]{Eric De Giuli}
\author[4,5]{Matthieu Wyart}
\author[2]{Yo\"el Forterre}
\author[1,*]{Jasna Bruji\'c}
\author[2,*]{Bloen Metzger}
\affil[1]{Center for Soft Matter Research, Department of Physics, New York University, New York, NY 10003, USA}
\affil[2]{Aix Marseille Univ., CNRS, IUSTI, Marseille 13453, France}
\affil[3]{Department of Physics, Toronto Metropolitan University, Toronto, ON M5B 2K3, Canada}
\affil[4]{Department of Physics and Astronomy, Johns Hopkins University, Baltimore, MD 21218, USA}
\affil[5]{Institute of Physics, \'Ecole Polytechnique F\'ed\'erale de Lausanne (EPFL), CH-1015 Lausanne, Switzerland}
\affil[*]{Corresponding authors: \href{mailto:jb2929@nyu.edu}{jb2929@nyu.edu}; \href{mailto:bloen.metzger@univ-amu.fr}{bloen.metzger@univ-amu.fr}}

\date{\normalsize Published in \emph{Proc.\ Natl.\ Acad.\ Sci.\ U.S.A.}\ \textbf{123}, e2611013123 (2026),
\href{https://doi.org/10.1073/pnas.2611013123}{doi:10.1073/pnas.2611013123}}

\begin{document}

\twocolumn[
\begin{@twocolumnfalse}
\maketitle

\section*{Abstract}
The flow of dense emulsions underlies applications from food and pharmaceutical processing to bioengineering, yet their rheology remains difficult to interpret under conventional volume-imposed conditions. Here, we change the control variable to osmotic-pressure using a recently developed instrument~--~the Capillarytron. This approach reveals a unified rheological structure where the osmotic pressure $\Pi$, by controlling droplet deformation, sets a \emph{pressure-dependent} jamming volume fraction. When expressed in terms of the distance to this jamming point, all rheological data~--~spanning both $\Pi$- and $\phi$-imposed measurements~--~collapse onto a single power-law divergence, akin to granular  suspensions. The resulting constitutive relations provide a predictive, parameter-free description of emulsion rheology across Newtonian, yielding, and shear-thinning regimes. Together with recent results on soft spheres, our findings point to a unifying paradigm: Soft amorphous materials~--~from soft spheres to emulsions and likely foams~--~obey the same \emph{hard rules} as granular suspensions, with softness entering through a pressure-dependent jamming point. This framework rationalizes Herschel–Bulkley rheology, assigns its parameters microscopic meaning, and opens perspectives on rigidity transitions in soft, deformable systems, including biological tissues.

\medskip
\noindent\textbf{Keywords:} Emulsion rheology $|$ Osmotic pressure $|$ Jamming transition $|$ Pressure-imposed rheology
\vspace{2em}
\end{@twocolumnfalse}
]

Dense emulsions are ubiquitous in cosmetic, food, and pharmaceutical formulations, and their rheology underlies processes ranging from 3D bioprinting to industrial food processing \cite{coussot2025fifty}. Because they combine droplet deformability with geometric crowding, emulsions have long served as model soft-amorphous materials for testing concepts spanning jamming, yielding, and nonequilibrium flow \cite{princen1987osmotic, mason1995elasticity}. More recently,  emulsions have also emerged as biomimetic analogs of living tissues, providing a model system in which elasticity, adhesion, and packing shape collective mechanics~\cite{jorjadze2011attractive,pontani2012biomimetic,guigue2025adhesion}. 

Despite this broad relevance, the rheology of dense emulsions still lacks a general microscopic framework and is predominantly described by the empirical Herschel–Bulkley (HB) law \cite{herschel1926}. Various theoretical approaches ranging from elastohydrodynamic lubrication models~\cite{meeker2004slip,seth2011micromechanical}, elastoplastic descriptions emphasizing avalanche-like rearrangements~\cite{hebraud1998mode,bocquet2009kinetic,nicolas2017deformation}, to jamming-based theories~\cite{wyart2012marginal,lin2014scaling}, have proposed derivations of the HB exponent from microscopic mechanisms. Nevertheless, a quantitative and predictive description of emulsion rheology, and its dependence on volume fraction, remains lacking. 

A central difficulty, absent in hard-sphere suspensions, arises from droplet deformability. Emulsions can be statically compressed beyond the hard-sphere jamming fraction $\phi_J$, thereby sustaining a finite yield stress~\cite{princen1987osmotic,mason1997osmotic,hohler2008osmotic,scheffold2013linear,maestro2013liquid,cohen2014rheology}. Under flow at fixed total volume, increasing shear stress $\tau$ raises the droplet normal stress~--~known as the osmotic or particle pressure  $\Pi$ \cite{princen1987osmotic,nott1994pressure,brady1997microstructure,morris1999curvilinear, hohler2008osmotic,deboeuf2009particle}~--~and induces additional droplet deformation~\cite{seth2011micromechanical}. This deformation effectively renormalizes the droplet size, leading to an apparent decrease of the emulsion volume fraction~\cite{berthier2009compressing,olsson2012herschel,chatte2018shear}. As a result, along a conventional rheogram, the effective distance to the jamming point evolves continuously with shear stress. This intrinsic coupling between shear stress, droplet deformation, and jamming obscures the interpretation of conventional flow curves and precludes the identification of the relevant control parameters.

Here we circumvent this problem by adopting \emph{pressure-imposed rheology}, in which the confining (osmotic) pressure $\Pi$ is fixed, while the droplet volume fraction $\phi$ is free to adjust when the shear rate $\dot{\gamma}$ is varied. This framework has recently unified the rheology of hard-sphere suspensions and deformable, elastic particles~\cite{boyer2011unifying,guazzelli2018rheology,kawasaki2015diverging,tapia2024rheology}, yet its applicability to emulsions remains unexplored. In contrast to soft solids such as hydrogels, emulsions are composed of liquid droplets whose deformability arises from interfacial tension rather than bulk elasticity. Under fixed $\Pi$, droplet deformation remains constant, thereby restoring a well-defined jamming point that no longer shifts with shear rate. Using a newly developed rheometer dedicated to colloidal systems \cite{etcheverry2023capillary}~--~the Capillarytron~--~we measure the rheology of dense emulsions under imposed pressure. We show that, as for elastic particles \cite{olsson2012herschel,kawasaki2015diverging,tapia2024rheology}, emulsion flow is entirely governed by the distance to a \emph{pressure-dependent} jamming volume fraction, $\phi_J'(\Pi)$, which encodes the level of droplet deformation and effectively removes all complications due to softness inherent to volume-imposed measurements. Recasting the rheology in terms of the distance to this boundary, $\Delta\phi = \phi_J'(\Pi) - \phi$ collapses all flow regimes~--~Newtonian, yield, and shear-thinning~--~onto predictive, parameter-free constitutive relations, which quantitatively reproduce independent volume-imposed measurements.

The previously developed pressure-dependent jamming framework for soft elastic spheres~\cite{olsson2012herschel,kawasaki2015diverging}, recently extended experimentally to hydrogel suspensions~\cite{tapia2024rheology}, finds here a direct realization in dense emulsions, whose deformability originates from interfacial tension rather than bulk elasticity. Taken together, these results support a unified description in which the rheology of soft amorphous materials~--~from soft spheres to emulsions~--~follows common granular suspension constitutive laws, with deformability entering primarily through a pressure-dependent jamming point. The breadth and precision of the Capillarytron measurements provide a particularly stringent experimental test of this framework and demonstrate its predictive power across a broad range of volume fractions and flow regimes. This pressure-organized framework rationalizes Herschel--Bulkley rheology, its dependence on volume fraction and provides a route toward understanding rigidity transitions in soft, deformable particulate matter, including biological tissues.

\begin{figure*}[t]\centering
\includegraphics[width=\textwidth]{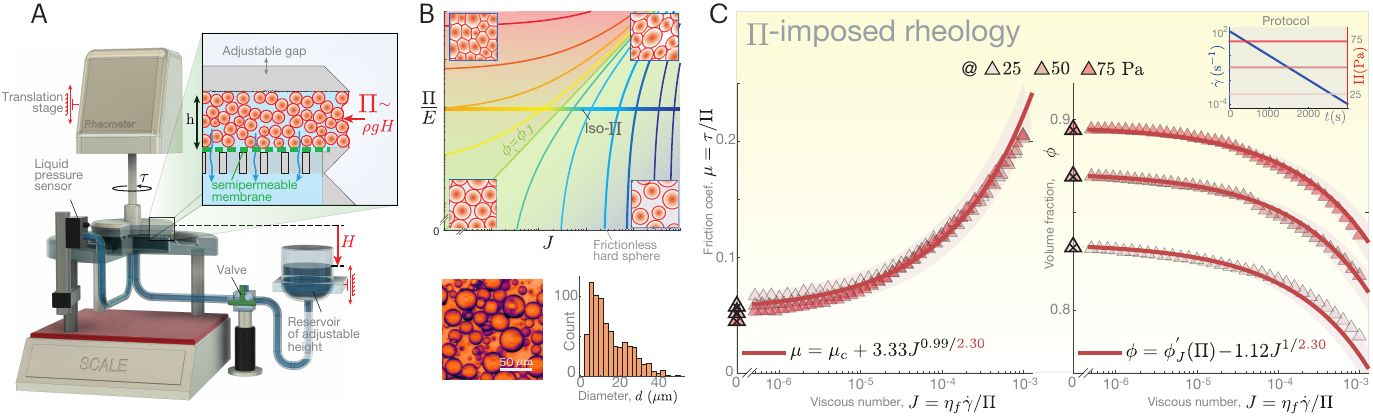}
\caption{ \textbf{A)} The Capillarytron~--~an osmotic-pressure-controlled rheometer. The emulsion is sheared in a plate–plate geometry while being hydraulically connected, through a semipermeable membrane, to a height-adjustable reservoir. Only the continuous phase flows through the membrane (blue arrows), allowing the droplet osmotic pressure $\Pi \sim \rho g H$ to be imposed by the capillary interface, while the volume fraction is free to vary. \textbf{B)} \textit{Top:} Control-parameter space of $\Pi$-imposed rheology. The viscous number $J=\eta_f\dot{\gamma}/\Pi$ and the normalized pressure $\Pi/E$ are externally controlled, while the emulsion volume fraction $\phi$ (color coded) adjusts freely. For deformable droplets, $\phi$ can exceed the hard-sphere jamming fraction $\phi_J$. \textit{Bottom:} Example microscope image of the silicone-oil droplets and their size distribution (mean diameter $d\simeq 15~\mu$m). \textbf{C)} $\Pi$-imposed rheological measurements. The friction coefficient $\mu=\tau/\Pi$ and the volume fraction $\phi$ are shown versus viscous number $J$ for three imposed pressures $\Pi$. Solid red curves show fits to Eqs.~\ref{eq:mu_J} and~\ref{eq:phi_J}. \textit{Inset:} Dynamic-ramp protocol: $\Pi$ is held fixed while $J$ is decreased quasi-steadily by lowering the shear rate $\dot{\gamma}$.}
\label{fig:fig1}
\end{figure*}

\section*{Osmotic-pressure-controlled rheology of emulsions}

The emulsion consists of silicone--oil droplets (viscosity $0.1~\mathrm{Pa\cdot s}$) dispersed and stabilized in an aqueous 8~mM SDS solution (Fig.~\ref{fig:fig1}B). Coarse mixing with a rotor--stator emulsifier produces droplets with mean diameter $d = 15~\mu\mathrm{m}$ and SD $\sigma_d = 9~\mu\mathrm{m}$, corresponding to a polydispersity $\mathrm{\epsilon} = \sigma_d / d = 0.73$~\cite{zhang2015structure,clusel2009granocentric}. The droplet ``elastic'' modulus, defined from the Laplace pressure, is $E = 2\gamma/d \approx 1.8~\mathrm{kPa}$, where the interfacial tension $\gamma = 13.6~\mathrm{mN/m}$ is measured using the pendant--drop method.

\emph{Osmotic--pressure--controlled rheology} is performed using a recently developed in-house instrument, the Capillarytron (Fig.~\ref{fig:fig1}A), described in detail in ref.~\cite{etcheverry2023capillary}. This device departs from conventional stress-, strain-, or volume-controlled rheometers by directly imposing the osmotic pressure $\Pi$ (the droplet confining pressure). The emulsion, held in a plate--plate geometry, is hydraulically coupled to a height-adjustable reservoir through a semipermeable membrane integrated into the bottom plate. Only the continuous phase flows through the membrane (blue arrows), allowing the steady state liquid pressure to be hydrostatically controlled. Lowering the reservoir increases the hydrostatic head and thus the osmotic pressure of the droplets, $\Pi \sim \rho g H$, imposed by the capillary interface. In the Capillarytron, $\Pi$ corresponds to the vertical normal stress confining the droplet assembly. Owing to stress anisotropy in the packed droplets, this stress can differ from the liquid pressure in the continuous phase (see ref.~\cite{etcheverry2023capillary} for details). To enforce no-slip boundary conditions, the top plate is roughened by placing a nylon mesh (250 $\mu$m) on top of the semipermeable membrane.  Shear is applied via the upper plate using a rheometer head (Anton Paar DSR~502). Any change in the imposed pressure $\Pi$ or shear rate $\dot{\gamma}$ causes the emulsion to compress or dilate; the resulting variations in sample volume are compensated by adjusting the gap $h$, from which the instantaneous volume fraction $\phi$ is obtained. Simultaneous measurements of shear stress $\tau$, and volume fraction $\phi$ under imposed pressure $\Pi$ and shear rate $\dot{\gamma}$ provide full access to the emulsion frictional rheology.

For dense suspensions of hard particles, the rheology expressed in pressure-imposed variables is fully described by the dimensionless viscous number $J=\eta_f\dot{\gamma}/\Pi$, together with the constitutive relations $[\mu(J), \phi(J)]$, where $\phi$ is the volume fraction, $\eta_f$ the viscosity of the suspending fluid, and $\mu$ the macroscopic friction coefficient. This effective friction should not be confused with interparticle solid friction, which is expected to be negligible for liquid droplets~\cite{boyer2011unifying,guazzelli2018rheology,etcheverry2023capillary}. Extending this framework to emulsions requires accounting for their deformability, which introduces an additional control parameter: $\Pi/E$, with $E=2\gamma/d$ the droplet ``elastic'' modulus. This extra dimensionless number broadens the rheological control--parameter space. As illustrated in Fig.~\ref{fig:fig1}B, the limit $\Pi/E \to 0$ (spherical droplets) should recover the hard, frictionless spheres limit, whereas increasing $\Pi/E$ drives the system into regimes where droplets appreciably deform. Operating at fixed pressure (moving along an iso-$\Pi$ line), while lowering the shear rate~--~i.e.\ decreasing $J = \eta_f \dot{\gamma}/\Pi$~--~forces the system to concentrate beyond the hard-sphere jamming limit $\phi_J$. The Capillarytron, which independently controls both $J$ and $\Pi/E$ thus provides direct experimental access to map the frictional rheology of soft emulsions, from undeformed droplets to strongly compressed, foam-like packings.

\paragraph*{Emulsions Follow the Laws of Granular Suspension Rheology.}
To characterize the flow behavior, we fix the imposed pressure $\Pi$ by setting the height of the reservoir and vary the viscous number $J$ by slowly and quasi-steadily decreasing the shear rate (protocol in Fig.~\ref{fig:fig1}C; see Materials and Methods). Fig.~\ref{fig:fig1}C displays the resulting friction coefficient $\mu = \tau/\Pi$ and volume fraction $\phi$ as a function of $J$ obtained under different pressures. Remarkably, the $\mu(J)$ curves are mostly unaffected by $\Pi$, indicating that the macroscopic friction coefficient can be considered pressure-insensitive. In contrast, the $\phi(J)$ curves reach progressively higher packing fractions as droplets become more compressed. These data are well described by simple power-law relations in $J$ %
\begin{align}
    \mu &= \mu_c + D\,B^{\gamma/\beta}\, J^{\gamma/\beta}, \label{eq:mu_J}\\
    \phi &= \phi_J' - B^{1/\beta}\, J^{1/\beta}, \label{eq:phi_J}
\end{align}
where $\mu_c \equiv \mu(J \to 0)$ is the quasistatic friction coefficient, $B$, $D$, $\beta$, and $\gamma$ are fitting parameters, and $\phi_J' \equiv \phi(J \to 0)$ is the $\Pi$-dependent jamming volume fraction. Interestingly, the \emph{dynamic} parts of these laws~--~the $J$-dependent terms~--~closely mirror those of hard-sphere granular suspensions. Both the reduced friction $\Delta\mu = \mu - \mu_c$ and the distance to jamming $\Delta\phi = \phi_J' - \phi$ follow a simple power law of $J$ with nearly identical exponents, $\gamma/\beta \approx 1/\beta \approx 1/2.30$ (\textit{SI Appendix}, section SI.1). The exponent value $\beta = 2.3$ lies in between the exponents known for frictionless ($\beta=2.8$) and frictional ($\beta=2$) hard-sphere~\cite{degiuli2015unified,perrin2019interparticle,etcheverry2023capillary}. The key difference between emulsions and hard-particle suspensions is thus not in the $J$-scaling, but in the pressure-dependent jamming transition.

\paragraph*{$\Pi$-dependent jamming point.}

\begin{figure}%
\centering
\includegraphics[width=1\linewidth]{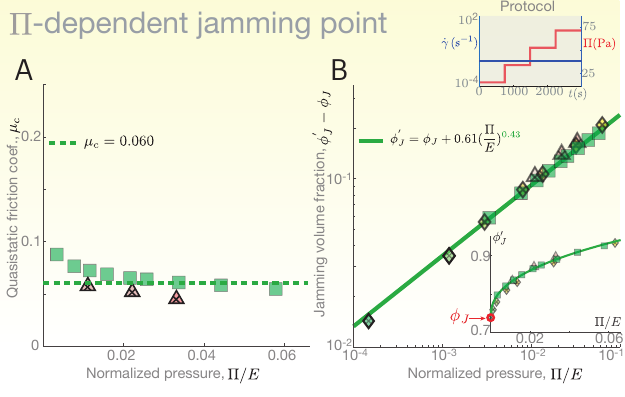}
\caption{\textbf{A)} Quasi-static friction coefficient $\mu_c$ and \textbf{B)} jamming volume fraction $\phi_J'$ versus imposed particle pressure $\Pi$.  Green squares: successive pressure steps while maintaining quasi-static shear ($J<10^{-6}$); Inset: experimental protocol.  Dashed line: mean quasi-static friction coefficient \eqref{eq:mu-c}. Solid line: \eqref{eq:phi_J_prime}.  
Red triangles: $\mu(J\!\to\!0,\Pi)$ and $\phi(J\!\to\!0,\Pi)$ from the $\Pi$-imposed dynamic ramps in Fig.~\ref{fig:fig1}C.  
Diamonds: yield pressures $\Pi_y=\tau_y/\mu_c$ obtained from the yield stress $\tau_y$ measured in the $\phi$-imposed protocol (Fig.~\ref{fig:scrut}A) and using $\mu_c=0.0602$.  Red circle: zero-pressure (hard-sphere) jamming point $\phi_J'(\Pi{=}0)=\phi_J=0.73\pm0.005$, obtained from the volume fraction at which a yield stress first appears in Fig.~\ref{fig:scrut}A.}
\label{fig:compressibility}
\end{figure}

To isolate the quasistatic response, we impose a very low shear rate that keeps the viscous number below $J < 10^{-6}$, i.e.\, in the immediate vicinity of jamming. We then probe the emulsion response to successive stepwise increases in $\Pi$ (protocol in Fig.~\ref{fig:compressibility}; see Materials and Methods). After each pressure step, the system relaxes to steady state, where we measure the dependence of $\mu_c \equiv \mu(J \to 0)$ and $\phi_J' \equiv \phi(J \to 0)$ on the imposed pressure.

As shown in Fig.~\ref{fig:compressibility}A, $\mu_c$ approaches $\sim 0.1$ as $\Pi \to 0$, which is the expected value for the geometric friction measured in frictionless hard spheres~\cite{PhysRevE.78.011307,clavaud2017revealing,perrin2019interparticle,etcheverry2023capillary}. With increasing $\Pi$, $\mu_c$ decreases modestly, reflecting the fact that droplet deformation reduces geometric interlocking and facilitates sliding. This trend aligns with previous observations in foams~\cite{lespiat2011jamming}. However, since this variation is small compared to the strong $J$-dependence of $\mu$, we neglect it and use a constant value
\begin{equation}
    \mu_c \approx 0.06,
    \label{eq:mu-c}
\end{equation}
which simplifies the analysis. The small value of the quasi-static friction coefficient, $\mu_c = \tau/\Pi \approx 0.06$ implies that the pressure in this regime exceeds shear stress by more than an order of magnitude. The dominant mechanical stress experienced by droplets is therefore the particle pressure, not the shear component.

In contrast to the weak pressure sensitivity of $\mu_c$, Fig.~\ref{fig:compressibility}B shows that the jamming volume fraction $\phi_J'$ exhibits a strong dependence on $\Pi$ (Fig.~\ref{fig:compressibility}B). As for elastic-particles \cite{kawasaki2015diverging,tapia2024rheology}, droplet deformability shifts the jamming transition, and the data collapse onto a simple compressibility law,
\begin{equation}
    \phi_J' = \phi_J + A\left(\frac{\Pi}{E}\right)^\alpha,
    \label{eq:phi_J_prime}
\end{equation}
where $\phi_J=\phi_J^{'} (\Pi=0)$ is the hard-sphere jamming fraction, $A= 0.61$ and $\alpha=0.43$ are fitting parameters. These parameters are tightly constrained because the intercept at zero pressure $\phi_J=0.73 \pm0.005$ is independently fixed by the volume fraction at which a yield stress first appears in $\phi$-imposed measurements (Fig.~\ref{fig:scrut}A). This result is further corroborated by the excellent agreement with the theoretical prediction of $\phi_J = 0.634 + 0.0278\,\epsilon + 0.196\,\epsilon^2 = 0.730$ based on the measured droplet polydispersity~\cite{zhang2015structure}. In addition to the robustness of the fit, we show that two other independent measurements give rise to the same compressibility law in Eq.~\ref{eq:phi_J_prime}: i) the asymptotic values $\phi(J\!\to\!0,\Pi)$ (red triangles) extracted from the dynamic ramps in Fig.~\ref{fig:fig1}C, and ii) the yield pressures $\Pi_y = \tau_y/\mu_c$ (diamonds)
obtained from the $\phi$-imposed yield stresses $\tau_y$ in Fig.~\ref{fig:scrut}A and using  $\mu_c = 0.06$. The quantitative agreement on the compressibility law across these three independent protocols demonstrates the robustness of Eq.~\ref{eq:phi_J_prime}, which captures a key property of emulsions~--~their pressure-dependent jamming volume fraction.

The pressure-imposed framework introduced above also gives insights to the compressibility of dense emulsions. Remarkably, although measured here under quasi-static shear, our data reproduce the empirical osmotic relation $\Pi/E \simeq K(\phi-\phi_J)^2/(1-\phi)^{1/2}$ with $K\simeq3.2$ previously reported for \emph{statically} compressed emulsions and foams~\cite{princen1987osmotic,mason1997osmotic,hohler2008osmotic,scheffold2013linear,maestro2013liquid,cohen2014rheology} (\textit{SI Appendix}, section SI.2 and Fig.~S2). The data are equally well captured by a simple power law $\Pi/E \sim (\phi-\phi_J)^{1/\alpha}$ (Eq. \ref{eq:phi_J_prime}), yielding $\alpha\simeq0.43$ for polydisperse emulsions and $\alpha\simeq0.5$ for more monodisperse systems (\textit{SI Appendix}, section SI.3). The slightly stiffer response at higher polydispersity is consistent with increased coordination in disordered packings~\cite{lacasse1996deformation,lacasse1996model}. Importantly, the pressure-dependent increase of $\phi_J'$  is also consistent with direct structural observations in compressed emulsions. Confocal imaging studies have shown that increasing compression progressively deforms droplets from nearly spherical to faceted polyhedral shapes, allowing packings to exceed the hard-sphere jamming fraction while remaining mechanically stable~\cite{jorjadze2013microscopic}, a behavior also addressed theoretically~\cite{schlegel2016local}. The microscopic origin of this scaling, however, remains unresolved and calls for further theoretical work.

\begin{figure}
\centering
\includegraphics[width=1\linewidth]{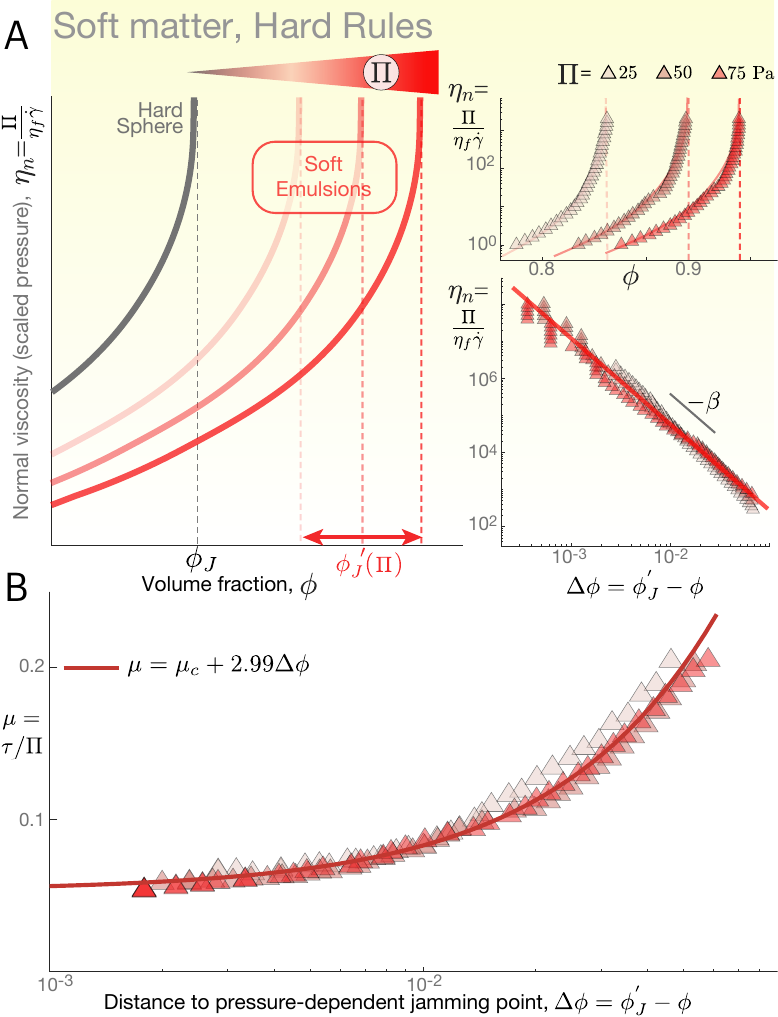}
\caption{\textbf{A)} Normalized pressure (normal viscosity) $\eta_n = \Pi/(\eta_f \dot{\gamma})$ versus $\phi$ according to Eq.~\ref{eq:pi}.  The pressure diverges as a simple power law of the distance to jamming, just as in hard spheres, but with a pressure-dependent jamming volume fraction $\phi_J'(\Pi)$.  \textit{Top inset:} $\eta_n=1/J$ versus $\phi$ obtained from the $\phi(J)$ data in Fig.~\ref{fig:fig1}C.   \textit{Bottom inset:} Same data versus distance to \emph{pressure-dependent} jamming point, $\Delta\phi = \phi_J' - \phi$, highlighting the collapse onto a single master curve. Red line: Eq.~\ref{eq:pi}. \textbf{B)} Emulsion friction coefficient $\mu$ versus $\Delta\phi\!=\!\phi_J'\!-\!\phi$. Red line: $\mu\!=\!\mu_c\!+\!2.99\,\Delta\phi$, obtained by canceling $J$ from Eqs.~\ref{eq:mu_J} and~\ref{eq:phi_J}.}
\label{fig:sumup}
\end{figure}

\section*{\boldmath Unifying $\Pi$- and $\phi$-imposed frameworks}

The $\Pi$-imposed constitutive framework (Eqs.~\ref{eq:mu_J}--\ref{eq:phi_J_prime}) reveals that emulsions behave like granular materials. Combining $J=\eta_f \dot{\gamma}/\Pi$ with Eq.~\ref{eq:phi_J}, directly recovers the standard volume-imposed expression for the normal viscosity \cite{brady1997microstructure,morris1999curvilinear, hohler2008osmotic,deboeuf2009particle},
\begin{equation}
\eta_n = \frac{\Pi}{\eta_f \dot{\gamma}} = \frac{B}{(\phi_J' - \phi)^\beta},
\label{eq:pi}
\end{equation}
which diverges as a power-law of the distance to jamming $\Delta\phi = \phi_J' - \phi$, as plotted in Fig.~\ref{fig:sumup}A. The key difference from hard spheres is that the jamming point is \emph{no longer} fixed: Droplet deformability shifts $\phi_J'(\Pi)$ upward with increasing pressure. This trend is directly observed from our pressure imposed data when plotting $\Pi/\eta_f \dot{\gamma}$ versus $\phi$ (Top Inset of Fig.~\ref{fig:sumup}A). Once this shift is accounted for by plotting $\Pi/\eta_f \dot{\gamma}$ versus $\Delta\phi$, all data collapse onto a single master curve (Bottom Inset), showing that emulsion rheology is entirely governed by the distance to a \emph{pressure-dependent} jamming point.

Such a simplicity arises here because the rheology is investigated and expressed under fixed particle pressure, for which $\phi_J'(\Pi)$ is \emph{constant}. By contrast, conventional volume-imposed measurements are far more difficult to interpret because i) increasing the applied stress continually shifts the location of jamming, and ii) given that $\mu(\Delta\phi)=\mu_c + D\Delta\phi$ is neither constant nor affine in $\Delta\phi$ (Fig.~\ref{fig:sumup}B), the shear stress $\tau=\mu\Pi$ is not a simple power law of $\Delta\phi$ (\textit{SI Appendix}, section SI.4).

In the next sections, we show that the $\Pi$-imposed flow rules can fully predict conventional volume-imposed rheology.

\paragraph*{$\Pi$-imposed flow rules predict $\phi$-imposed measurements.}

\begin{figure*}
\centering
\includegraphics[width=1\linewidth]{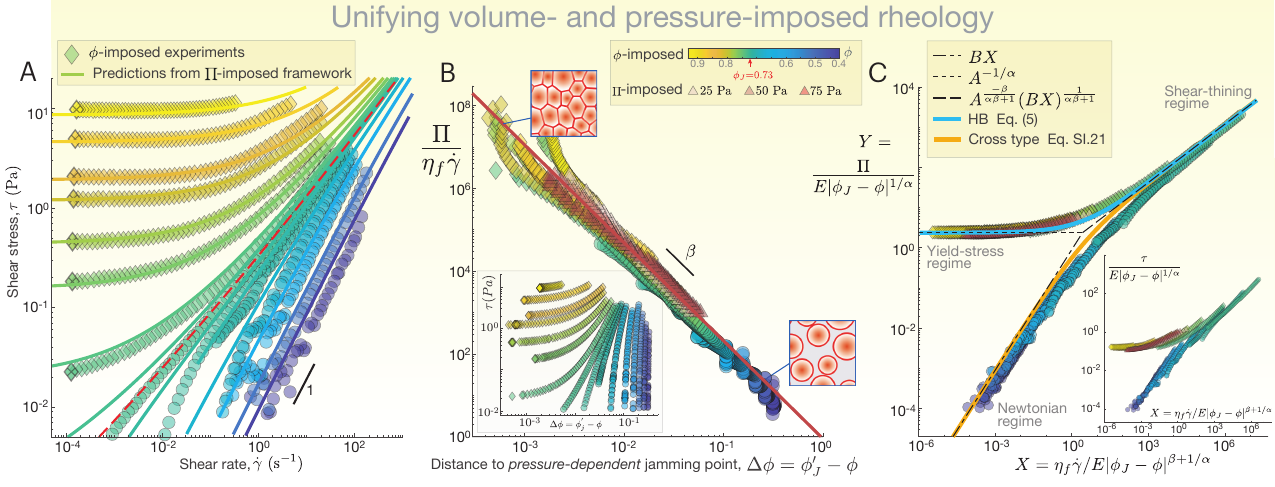}
\caption{\textbf{A)} Conventional $\phi$-imposed rheograms~--~shear stress $\tau$ versus shear rate $\dot{\gamma}$~--~for different imposed volume fractions $\phi \in [0.40-0.93]$. 
Symbols: $\phi$-imposed measurements below (circles) and above (diamonds) $\phi_J$. Solid lines: predictions from $\Pi$-imposed framework (Eqs.~\ref{eq:mu_J}--\ref{eq:phi_J_prime}). Red dashed-line: $\phi=\phi_J$, the hard sphere jamming volume fraction. Bold-diamonds: yield shear stress $\tau_y$ yielding compressibility law in Fig. \ref{fig:compressibility}.
\textbf{B)} Rescaled pressure $\Pi/\eta_f\dot{\gamma}$ versus distance to \emph{pressure-dependent} jamming point  $\Delta \phi=\phi_J'-\phi$. Circles and diamonds: $\phi-$imposed data from (A) with volume fractions $\pm 20\%$ above and below $\phi_J$. Red line: \eqref{eq:pi}. Inset: mapping of $\tau(\dot{\gamma})$ from (A) on $\tau(\Delta \phi)$ based on $\Pi$ imposed framework.  \textbf{C)} Scaled pressure  $Y = \Pi/E|\phi_J - \phi|^{1/\alpha}$ versus scaled shear rate $X = \eta_f \dot{\gamma}/E|\phi_J - \phi|^{\beta + 1/\alpha}$ reflecting the emulsion behavior above and below $\phi_J$. Dashed lines: Newtonian, yield and shear thinning asymptotic regimes. Blue line: HB relation \eqref{eq:HBforP}. Inset: Corresponding rescaling for $\tau$. All parameters are fixed  (within uncertainties) from constitutive $\Pi$-imposed relations (Eqs.~\ref{eq:mu_J}--\ref{eq:phi_J_prime}).}
\label{fig:scrut}
\end{figure*}

By closing the valve (Fig.~\ref{fig:fig1}A), the Capillarytron operates in standard $\phi$-imposed mode. We then measure classical rheograms~--~shear stress $\tau$ versus shear rate $\dot{\gamma}$~--~over a broad range of volume fractions $\phi \in [0.40{-}0.90]$ (Materials and Methods). As shown in Fig.~\ref{fig:scrut}A, the emulsion displays the canonical sequence of flow regimes: a Newtonian response ($\tau \sim \dot{\gamma}$) for $\phi < \phi_J$ and low $\dot{\gamma}$, the emergence of a yield shear-stress $\tau_y$ for $\phi > \phi_J$ and low $\dot{\gamma}$, and shear thinning ($\tau \sim \dot{\gamma}^n$ with $n<1$) at large $\dot{\gamma}$.

The key result in Fig.~\ref{fig:scrut}A is that the constitutive relations (Eqs.~\ref{eq:mu_J}--\ref{eq:phi_J_prime}), determined independently from the $\Pi$-imposed measurements in Figs.~\ref{fig:fig1}C and~\ref{fig:compressibility}, quantitatively predict the $\phi$-imposed rheograms (\textit{SI Appendix}, section SI.5.A) without any free adjustable parameters (\textit{SI Appendix}, section SI.6). Remarkably, although these flow rules are established in the dense regime $\phi>\phi_J$, they accurately reproduce the rheology down to $\phi=0.40$, deep in the dilute Newtonian regime. The pressure-imposed framework therefore captures emulsion rheology and its volume-fraction dependence across all flow regimes.

\paragraph*{$\Pi$ is the central variable collapsing all rheological data.}

Volume-imposed measurements provide only $\tau(\dot{\gamma})$ curves (Fig.~\ref{fig:scrut}A). These raw rheograms are difficult to interpret because, along an iso-$\phi$ curve, each point corresponds to a different droplet pressure $\Pi=\tau/\mu$ and therefore to a different location of jamming $\phi_J'(\Pi)$. The $\Pi$-imposed relations (Eqs.~\ref{eq:mu_J}--\ref{eq:phi_J_prime})  provide a unique mapping  between the measured stress $\tau$ and the distance to jamming $\Delta\phi = \phi_J' - \phi$, as illustrated in the Inset of Fig.~\ref{fig:scrut}B (\textit{SI Appendix}, section SI.5A). Using this inversion, the $\phi$-imposed data can be reexpressed in terms of the granular scaling 
$\Pi/\eta_f \dot{\gamma}
    = \tau/\eta_f \dot{\gamma}\mu(\Delta\phi)
    = B/\Delta\phi^\beta$ (Eq.~\ref{eq:pi} and \textit{SI Appendix}, section SI.5B).
As shown in Fig.~\ref{fig:scrut}B, \emph{all} $\phi$-imposed measurements collapse onto the same master curve previously identified under $\Pi$-imposed conditions (\eqref{eq:pi} and Fig.~\ref{fig:sumup}). The osmotic pressure $\Pi$ thus unifies the rheology obtained under both $\Pi$-imposed and $\phi$-imposed protocols. The collapse shows no detectable asymmetry between data above and below the hard-sphere jamming point $\phi_J$. This symmetry is all the more remarkable given that the analysis spans a broad range of volume fractions~--~far beyond the narrow jamming vicinity $\phi_J \pm 0.5\%$, as in ref.~\cite{olsson2012herschel}~--~covering more than $\pm 20\%$ around $\phi_J$, from dilute Newtonian emulsions to highly compressed, facetted droplet packings. 

\paragraph*{Osmotic pressure $\Pi$, not shear stress $\tau$, obeys Herschel--Bulkley rheology.} Following the approach of refs.~\cite{olsson2007critical,olsson2012herschel,kawasaki2015diverging,tapia2024rheology}, the $\Pi$-imposed constitutive laws (Eqs.~\ref{eq:mu_J}--\ref{eq:phi_J_prime}) allow us to derive the following scaling function (\textit{SI Appendix}, section SI.5C): 
\begin{align}
    Y = \frac{B X}{\left( \pm 1 + A Y^\alpha \right)^\beta}, 
    \label{eq:scaling2}
\end{align}
\noindent where the shear rate and the pressure
\[
    X = \frac{\eta_f \dot{\gamma}}{E|\phi_J - \phi|^{\beta + 1/\alpha}}, 
    \quad 
    Y = \frac{\Pi}{E|\phi_J - \phi|^{1/\alpha}},
\]
are this time rescaled by distance to the hard-sphere jamming limit $|\phi_J - \phi|$. Plotting $Y$ versus $X$ in Fig.~\ref{fig:scrut}C collapses \emph{all} data~--~both $\phi$-imposed and $\Pi$-imposed~--~onto two branches  reflecting the emulsion behavior above and below $\phi_J$. Eq.~\ref{eq:scaling2} further yields three asymptotic limits (Newtonian, yield-stress, and shear-thinning), shown as dashed lines in Fig.~\ref{fig:scrut}C (see derivation in \textit{SI Appendix}, SI.5C). For $\phi>\phi_J$, summing the yield and shear-thinning asymptotic contributions shows that the osmotic pressure $\Pi$ obeys the Herschel--Bulkley (HB) law~\cite{herschel1926konsistenzmessungen}
\begin{equation}
\frac{\Pi}{E}_{|_{\phi>\phi_J}} =  \left(\frac{\phi - \phi_J}{A}\right)^{1/\alpha}+A^{-\frac{\beta}{\alpha\beta + 1}}(B \eta_f \dot{\gamma}/E)^{\frac{1}{\alpha\beta  +1}},
    \label{eq:HBforP}
\end{equation}
represented by the blue line in Fig.~\ref{fig:scrut}C. Similarly, we obtain an explicit expression of $\Pi$ for $\phi<\phi_J$ (Fig. \ref{fig:scrut}C, orange line) as detailed in \textit{SI Appendix}, section SI.7. Eq.~\ref{eq:HBforP} reveals the physical origin of long-standing empirical HB fit parameters: the HB yield term $\left(\phi - \phi_J/A\right)^{1/\alpha}$~--~set by the compressibility law \eqref{eq:phi_J_prime}~--~and the HB exponent $n^\Pi_{\rm{exp}}=1/(\alpha \beta+1)$, shows that shear thinning in emulsions arises from the interplay of both the granular scaling exponent $\beta$ and the compressibility law exponent $\alpha$. 
Interestingly, the HB exponent directly follows from Eqs.~\ref{eq:phi_J} and \ref{eq:phi_J_prime}, which in the shear-thinning limit $\phi \to \phi_J$, yield $\Pi/E \sim J^{1/\alpha \beta}$. Solving for $\Pi(\dot{\gamma})$ with $J=\eta_f \dot{\gamma}/\Pi$, one recovers $\Pi \sim \dot{\gamma}^{1/(\alpha\beta  +1)} \approx \dot{\gamma}^{1/2}$.  The same expression for the shear-thinning exponent, $n^\Pi_{\rm{exp}}=1/(\alpha \beta+1)$, has been reported in numerical simulations~\cite{kawasaki2015diverging} and pressure-imposed experiments on soft elastic spheres~\cite{tapia2024rheology}, but with different values $n^\Pi_{\rm{exp}}\approx 0.3-0.39$, owing to system-specific compressibility and granular scaling exponents. Our framework and experimental data also show that the shear stress $\tau=\mu(\Delta\phi)\,\Pi$ cannot strictly be written in a simple scaling form. This is directly reflected in the poor collapse of $\tau$ observed in the Inset of Fig.~\ref{fig:scrut}C. Together, these results demonstrate that i) the osmotic pressure $\Pi$, not the shear stress $\tau$, is the natural Herschel--Bulkley variable governing emulsion rheology, and ii) the HB exponent follows directly from the granular scaling exponents.  

Within this framework, an approximate expression for $\tau$ can be derived (\textit{SI Appendix}, section SI.7):
\begin{equation}
\frac{\tau}{E}_{|_{\phi>\phi_J}} \approx \mu_c \left(\frac{\phi - \phi_J}{A}\right)^{1/\alpha} + A^{-\frac{\beta^{\prime}}{\alpha\beta^{\prime} + 1}}\left(B \eta_f \dot{\gamma}/E\right)^{\frac{1}{\alpha\beta^{\prime} +1}},
\label{eqSI:HB for tau approx}
\end{equation}
with $\beta'=1.4$. This expression yields a shear-thinning exponent for $\tau$ of $n^\tau_{\rm exp}=1/(\alpha\beta'+1)$. Interestingly, elastoplastic theories of amorphous solids, which apply in the regime $\tau-\tau_c \ll \tau_c$, predict a scaling of the form $\tau - \tau_c \sim \dot{\gamma}^{n_{\rm e}}$~\cite{Argon1979,FalkLanger1998,Baret2002,Picard2004,MaloneyLemaitre2004,LemaitreCaroli2009,Dahmen2011,lin2014scaling,Salerno2013,Talamali2011,Nicolas2018}. Although no consensus has emerged on the universality of the exponent $n_{\rm e}$, reported values $n_{\rm e}\in[0.59,0.66]$ (typically expected at low $\dot{\gamma}$)~\cite{Nicolas2018,lin2018microscopic} are remarkably close to the value $n^\tau_{\rm exp}=1/(\alpha\beta'+1)\simeq0.62$ obtained here at large $\dot{\gamma}$. The proximity of these exponents, despite the distinct physical regimes they correspond to, may explain why a single effective exponent~--~as in Eq. \ref{eq:mu_J}~--~provides an adequate description of $\mu(J)$ over the full range explored, and why no clear scaling asymmetry is observed in Fig.~\ref{fig:scrut}B across $\phi_J$. 

Discriminating between these regimes would require more sensitive probes, such as stop-flow experiments~\cite{lerner2012unified,hedstrom2024considerations,vinutha2024memory} or measurements of dynamical heterogeneities and nonlocal effects~\cite{Goyon2008Spatial,Kamrin2019NonlocalReview,perrin2021nonlocal,Ovarlez2012Nonlocal,Gueudre2016Scaling}. More generally, the constitutive laws developed here are restricted to steady-state rheology and do not describe transient, history-dependent, or multidirectional flows \cite{blanc2026disorder}, for which additional structural variables are known to play an important role. Likewise, nonlocal effects in confined flows may require supplementary constitutive ingredients beyond the bulk steady-state framework developed here.

\section*{Conclusion} 
The $\Pi$-imposed rheology experiments establish a robust, predictive, and parameter-free framework (Eqs.~\ref{eq:mu_J}--\ref{eq:phi_J_prime}) that captures the rheology of emulsions across their Newtonian, yielding, and shear-thinning regimes. Relative to hard spheres, droplet softness primarily manifests through an upward shift of the jamming point, $\phi_J^{\prime}(\Pi)$, as droplets are increasingly deformed under pressure. Expressing the rheology in terms of the distance to this pressure-dependent jamming point, $\Delta\phi=\phi_J^{\prime}-\phi$, removes the dominant, nonuniversal effects of softness. In this representation, $\phi_J^{\prime}(\Pi)$ absorbs the additional length and energy scales associated with droplet deformability, leaving a single geometric control variable~--~$\Delta\phi$~--~that governs the flow and yields a universal, granular-like power-law scaling for the pressure (Eq.~\ref{eq:pi}).

This framework is consistent with earlier numerical studies of soft spheres by Olsson and Teitel~\cite{olsson2012herschel}, Kawasaki et al.~\cite{kawasaki2015diverging}, and with recent pressure-imposed experiments on soft hydrogels by Tapia et al.~\cite{tapia2024rheology}. Together, these studies support a unified description in which the rheology of soft amorphous materials follows the common constitutive laws of granular suspensions, with deformability entering primarily through a pressure-dependent jamming point. The present work extends this picture to dense emulsions and demonstrates experimentally that the resulting pressure-controlled description remains predictive across Newtonian, yielding, and shear-thinning regimes. This framework gives physical meaning to the parameters previously used empirically in Herschel--Bulkley fits and provides a natural link between jamming and elastoplastic theories by explaining the emergence of a single effective exponent across flow regimes. Interestingly, the concept of a pressure-dependent jamming volume fraction extends beyond soft shear-thinning systems~\cite{Gueudre2016Scaling}. It also underlies discontinuous shear thickening in dense suspensions, where increasing particle pressure, by activating interparticle frictional contacts, instead lowers the jamming volume fraction and can result in a dramatic viscosity increase~\cite{chatte2018shear,blanc2026disorder,lenne2022sculpting,petridou2021rigidity}. In this sense, both shear thickening and soft-shear-thinning rheology can be viewed within a common framework in which pressure controls the location of the jamming transition, albeit through distinct microscopic mechanisms.

The overarching message is that particle pressure~--~rather than shear stress~--~is the central organizing variable of emulsion rheology. Progress in the field has thus long been limited by an implicit experimental bias: Conventional rheometers measure or impose shear stress, whereas pressure~--~the quantity that controls particle deformation and the location of jamming~--~is neither controlled nor directly accessible. Identifying pressure as an organizing variable may have implications beyond passive emulsions, particularly for soft, deformable amorphous materials such as biological tissues. Tissue mechanics has recently adopted the language of jamming to interpret rigidity transitions during morphogenesis~\cite{lenne2022sculpting,petridou2021rigidity}. In many current models, including vertex and self-propelled vertex models, rigidity transitions at fixed volume fraction are described primarily through changes in a dimensionless cell-shape index~\cite{bi2015density,bi2016motility}. Our results demonstrate that, in emulsions, rigidity transitions can be driven simply by varying osmotic pressure. Whether a similar pressure-controlled mechanism contributes to rigidity transitions in biological tissues remains an open question worthy of further investigation.

The framework also appears robust to changes in droplet-size distribution: A lower-polydispersity soybean-oil emulsion (\textit{SI Appendix}, section SI.3) exhibits a markedly lower jamming fraction ($\phi_J=0.66$) but similar pressure-imposed constitutive exponents.  This suggests that, within the range of polydispersities explored here, polydispersity primarily affects the location of jamming rather than the form of the pressure-controlled constitutive laws. We have not explored the role of the droplet–continuous-phase viscosity ratio. While this additional dimensionless parameter is not expected to alter the overall framework, it may affect the value of the scaling exponent $\beta$; in the limit of large viscosity ratio, one would expect recovery of the frictionless hard-sphere behavior with $\beta \approx 2.8$~\cite{degiuli2015unified,perrin2019interparticle,etcheverry2023capillary}. 

More broadly, the $\Pi$-imposed methodology and constitutive laws provide a basis for extending this approach to other systems, including adhesive or Brownian emulsions and foams. In such systems, additional stress scales—arising from adhesion or activity—may shift $\phi_J'(\Pi)$ or renormalize the constitutive relations. Determining how these ingredients integrate into a pressure-controlled jamming framework is particularly relevant for biological tissues, where active stresses, adhesion-mediated contacts, and pressure  jointly govern rigidity, and morphogenetic flows~\cite{levental2007soft, matoz2017nonlinear, tlili2020migrating, gsell2025marangoni}.  Exploring these directions may reveal whether pressure remains the primary organizing variable~--~or whether new control parameters emerge when activity or adhesion compete with osmotic constraints.

\section*{Materials and Methods}
\textbf{Materials} Silicone oil emulsions are prepared at a volume fraction of 30\% using a high-speed shearing emulsifier. Briefly, silicone oil 47V100 (density = 0.965~g/cm$^3$, viscosity = 0.1~Pa$\cdot$s) and an 8~mM SDS/water solution are mixed in a 3:7 volume ratio, then emulsified at 2,500~rpm for 10~min followed by 3,000~rpm for 20~min.  Soybean oil emulsions (\textit{SI Appendix}) are fabricated via membrane emulsification using 2.1-$\mu\mathrm{m}$-pore membrane (SPG Technology, Miyazaki, Japan), with an aqueous solution containing 20~mM phosphate buffer and 6~mM Tween 20 as the continuous phase. The emulsion is subsequently washed three times and finally stabilized with an 8~mM SDS/Water solution. The volume fraction of the emulsion is calibrated from the hard-sphere jamming limit estimated from the droplet polydispersity. Membrane filters (MCE, 3.0~$\mu$m pore size, 47~mm diameter, hydrophilic) are purchased from Sigma-Aldrich.

\textbf{\boldmath Frictional rheology: dynamic ramp protocol} From an initial pressure-imposed stationary state, a slowly descending ramp of viscous number $J$ is applied under
constant osmotic pressure $\Pi$.  This is implemented through a logarithmic reduction of the rheometer head rotation speed from $1~\mathrm{rps}$ down to $10^{-6}~\mathrm{rps}$, while maintaining the height of the reservoir $H$ constant. As the shear rate decreases under constant pressure, the volume
fraction gradually increases. As the scale reads a mass $m=S(\Pi-\rho g H)$/g, where $S$ is the surface of the rheometer plates and $\rho$ the continuous phase density, we have direct access the particle pressure. The gap $h$ is constantly adjusted in order to keep $\Pi$ constant (which is equivalent to maintaining the global curvature of the capillary interface constant, see ref.~\cite{etcheverry2023capillary} for more details). Variations of $h$ provides variations in the sample volume fraction.

\textbf{\boldmath $\Pi$-dependent jamming point: quasi-static pressure steps} Here, we impose a very low and fixed shear rate, that keeps the viscous number below $J < 10^{-6}$, i.e.\ in the immediate vicinity of jamming, while the pressure is increased in successive steps. After each step-increase in pressure, obtained by lowering the reservoir, we let the system relax to its steady state volume fraction $\phi_J'(\Pi)$. 

\textbf{\boldmath Volume-imposed rheology} The valve is closed to turn the Capillarytron in volume-imposed mode. Measurements are performed after a preshear at $1~\mathrm{rps}$ for 50~s to ensure homogeneity, followed by a quasi-steady ramp down from $1~\mathrm{rps}$ to $10^{-6}~\mathrm{rps}$, during which data are collected.  The Capillarytron allows us to withdraw fluid from the sample \emph{in situ}, simply by opening the valve and lowering the reservoir. Measurements are then repeated at different volume fractions. 

\textbf{Data, Materials, and Software Availability} Data have been deposited in Zenodo (DOI: \href{https://doi.org/10.5281/zenodo.19250640}{10.5281/zenodo.19250640})~\cite{chen2026data}.

\textbf{Acknowledgments} We thank Pascale Aussillous, Elisabeth Guazzelli, Emanuela Del Gado, Frank Scheffold, Jacques Fattaccioli, and Lea-Laetitia Pontani for discussions. J.B. is supported by the National Science Foundation, Division of Materials Research grant No. 2105255 and the Swiss NSF through Grant No. 10000141. B.M. is supported by Agence Nationale de la Recherche ScienceFriction (ANR-18-CE30-0024).

\bibliographystyle{unsrt}
\bibliography{references}

\onecolumn
\let\section\SIsection
\setcounter{figure}{0}
\setcounter{table}{0}
\setcounter{equation}{0}
\setcounter{subsection}{0}
\renewcommand{\thefigure}{S\arabic{figure}}
\renewcommand{\thetable}{S\arabic{table}}
\renewcommand{\theequation}{S\arabic{equation}}
\renewcommand{\thesubsection}{\Alph{subsection}}
\renewcommand{\theHfigure}{S\arabic{figure}}
\renewcommand{\theHtable}{S\arabic{table}}
\renewcommand{\theHequation}{S\arabic{equation}}
\renewcommand{\theHsubsection}{SI.\Alph{subsection}}
\makeatletter
\renewcommand{\@seccntformat}[1]{\csname the#1\endcsname.\quad}
\makeatother

\begin{center}
{\LARGE\bfseries Supporting Information}\par\medskip
{\large Soft Matter, Hard Rules: Emulsions Follow the Laws of Granular Suspension Rheology}\par\medskip
Wenjun Chen, Eric De Giuli, Matthieu Wyart, Yo\"el Forterre, Jasna Bruji\'c, and Bloen Metzger
\end{center}
\bigskip

\noindent SI.1 Reduced friction coefficient $\Delta \mu=\mu-\mu_c$ and distance to jamming $\Delta \phi= \phi_J' - \phi$ \\
SI.2 Compressibility law\\
SI.3 Effect of polydispersity\\
SI.4 Osmotic pressure $\Pi$, not shear stress $\tau$, is the central variable \\
SI.5 Derivation of the $\phi$-imposed rheology and scaling analyses\\
SI.6 Parameters of the pressure imposed framework\\
SI.7 Osmotic pressure $\Pi$, not shear stress $\tau$, obeys Herschel-Bulkley rheology \\

\section*{SI.1 Reduced friction coefficient $\Delta \mu=\mu-\mu_c$ and distance to jamming $\Delta \phi= \phi_J' - \phi$.}

\begin{figure}[tbhp]
\centering
\includegraphics[width=0.55\textwidth]{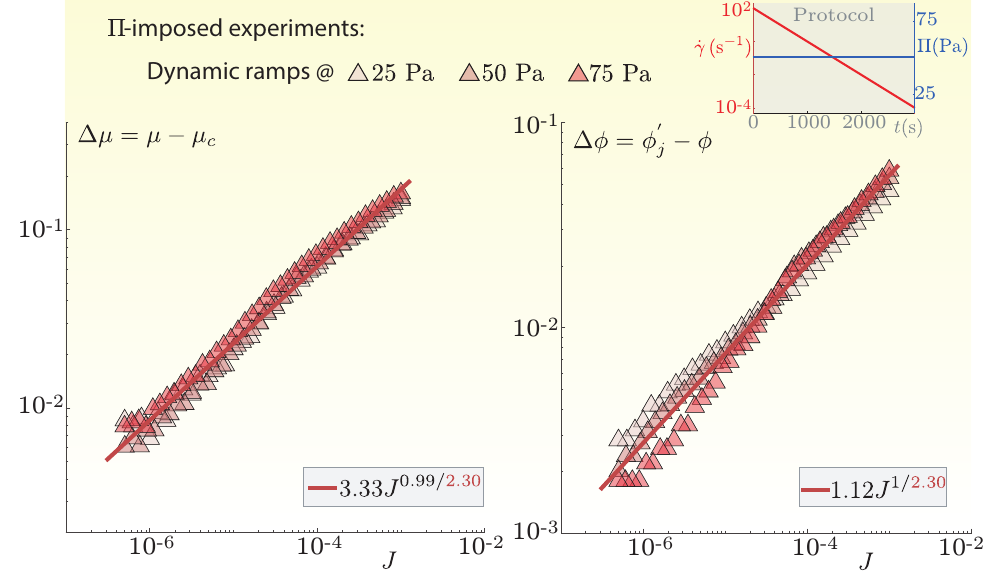}
\caption{Data are from the $\Pi$-imposed measurements in Fig.~1 of the main text. Reduced friction coefficient $\Delta \mu=\mu-\mu_c$ (left) and the distance to jamming $\Delta \phi= \phi_J' - \phi$ (right) versus viscous number $J=\eta_f \dot{\gamma}/\Pi$ for different imposed pressures $\Pi$. Solid lines:  $\Delta \mu \sim J^{0.99/2.30}$ and $\Delta \phi \sim J^{1/2.30}$, respectively. Inset: measurement protocol, $\Pi$ is fixed while $J=\eta_f \dot{\gamma}/\Pi$ is varied through a quasi-steady, slow decrease of shear rate $\dot{\gamma}$.}
\end{figure}

\section*{SI.2 Compressibility law}

The pressure-imposed framework gives new insights to the compressibility of dense emulsions. Remarkably, although measured here under quasi-static shear, our data reproduce the empirical osmotic relation 
\begin{equation}
\tilde{\Pi} \equiv \frac{\Pi}{2\gamma/d} \simeq K \frac{(\phi-\phi_J)^2}{(1-\phi)^{1/2}},
\label{SI-eqSI:model2}
\end{equation}
with $K\simeq3.2$ previously reported for \emph{statically} compressed emulsions and foams~\cite{princen1987osmotic,mason1997osmotic,hohler2008osmotic,scheffold2013linear,maestro2013liquid,cohen2014rheology} (SI Fig.~S2). The data are equally well captured by a simple power law $\Pi/E \sim (\phi-\phi_J)^{1/\alpha}$ (Eq.~4 in main text), yielding $\alpha\simeq0.43$ for polydisperse emulsions and $\alpha\simeq0.5$ for more monodisperse systems (see SI.3 below). 
This weak dependence on polydispersity suggests that droplet size distribution plays only a secondary role. The observed trend~--~larger polydispersity yielding slightly smaller $\alpha$, i.e. a stiffer response~--~is consistent with theoretical arguments linking increased coordination number in polydisperse packings to enhanced effective stiffness~\cite{lacasse1996deformation,lacasse1996model}. Although previous works have fitted compressibility data using more elaborate phenomenological expressions~\cite{lacasse1996model,scheffold2013linear,cohen2014rheology}, the present results show that a simple power law provides an equally accurate description across a wide range of systems. The microscopic origin of this scaling, however, remains unresolved and calls for further theoretical investigation.
\begin{figure}[tbhp]
\centering
\includegraphics[width=0.35\textwidth]{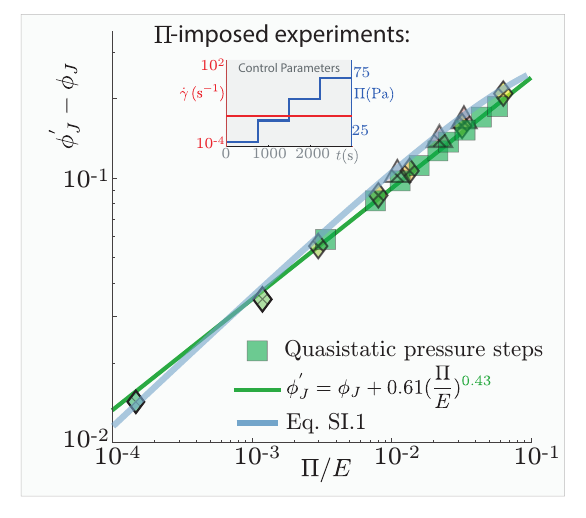}
\caption{Compressibility law. Pressure-dependent shift of the jamming volume fraction, $\phi_J' - \phi_J$, as a function of the osmotic pressure $\Pi$. Our measurements (green curve and symbols) are in quantitative agreement with the theoretical prediction given by Eq.~\ref{SI-eqSI:model2} (blue line). Green squares: successive pressure  steps (protocol shown in the inset). Red triangles: $\mu(J\!\to\!0,\Pi)$ and $\phi(J\!\to\!0,\Pi)$ from the $\Pi$-imposed dynamic ramps in Fig.~1.C.  
Diamonds: yield pressures $\Pi_y=\tau_y/\mu_c$ obtained from the yield stress $\tau_y$ measured in the $\phi$-imposed protocol (Fig.~4.A) and using $\mu_c=0.0602$. .}
\label{SI-fig:compressibility-theory}
\end{figure}

\begin{figure}[htbp]
\centering
\includegraphics[width=1\textwidth]{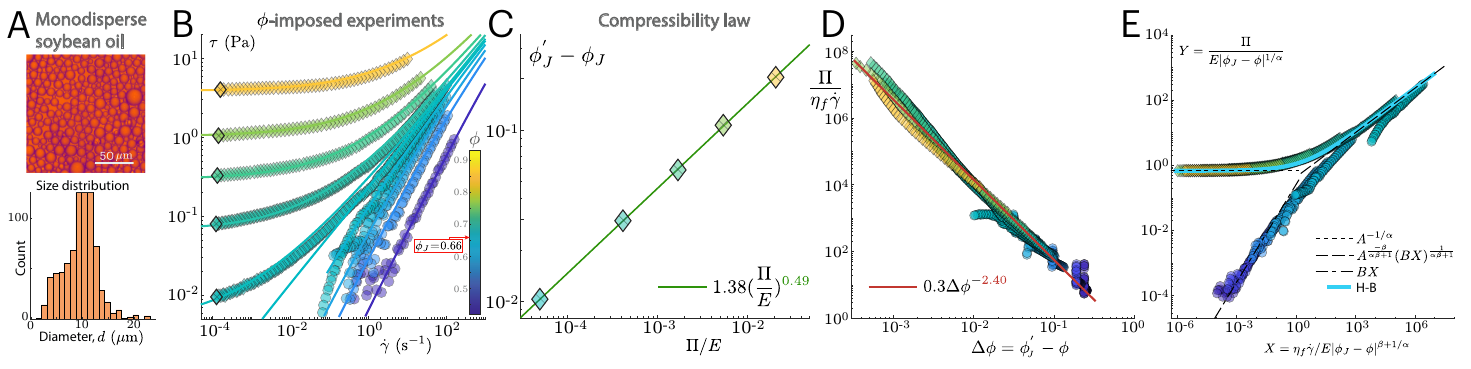}
\caption{Monodisperse emulsion.
(A) Optical micrograph of the droplets and corresponding size distribution obtained from $N=877$ droplets. The droplets have a mean diameter $d = 9.6~\mu$m and a standard deviation $\sigma_d = 3.1~\mu$m. 
(B) $\phi$-imposed rheology: shear stress $\tau$ versus shear rate $\dot{\gamma}$ for volume fractions (color-coded) ranging from $43\%$ to $86\%$ (circles: $\phi<\phi_J$; diamonds: $\phi>\phi_J$). A finite yield stress emerges above $\phi_J \approx 66 \pm 0.5\%$. Solid lines show predictions of the constitutive laws using the parameters obtained from the collapse in (D). 
(C) Compressibility relation obtained from the yield stress in (B), using $\Pi=\tau_y/\mu_c$ with $\mu_c = 0.0602$ and $\phi_J = 66 \pm 0.5\%$. 
(D) Rescaled pressure $\Pi/(\eta_f\dot{\gamma})$ plotted as a function of the distance to jamming $\Delta\phi=\phi_J'-\phi$. 
(E) Scaling collapse of $\Pi$ versus $\dot{\gamma}$ using the exponents $1/\alpha$ and $\beta+1/\alpha$ obtained from the collapse in (D).}
\label{SI-fig:monodisperse}
\end{figure}

\section*{SI.3 Effect of polydispersity}

To assess the influence of droplet polydispersity, we prepared a soybean-oil emulsion by \emph{membrane emulsification} using a 2.1-$\mu$m-pore SPG membrane (SPG Technology, Japan). The continuous phase consisted of an aqueous solution of 20 mM phosphate buffer and 6 mM Tween 20. The emulsion was washed three times and finally stabilized in an 8 mM SDS aqueous solution. As shown in Fig.~\ref{SI-fig:monodisperse}.A, the droplets have a mean diameter $d = 9.6~\mu$m and a standard deviation $\sigma_d = 3.1~\mu$m, corresponding to a polydispersity $\epsilon = \sigma_d/d = 32\%$. Using the empirical relation $\phi_J = 0.634 + 0.0278\epsilon + 0.196\epsilon^2$ from Ref.~\cite{zhang2015structure}, we estimate the hard-sphere jamming volume fraction as $\phi_J \simeq 0.66$, which is used to calibrate the emulsion volume fraction. Since soybean-oil droplets have an interfacial tension comparable to silicone oil, we take $\gamma = 13.6$ mN/m, yielding a droplet Young’s modulus $E = 2\gamma/d = 2.8 \times 10^{3}$~Pa.

Using the Capillarytron in its $\phi$-imposed configuration, we measured flow curves $\tau(\dot{\gamma})$ over volume fractions $0.43 \le \phi \le 0.86$ (Fig.~\ref{SI-fig:monodisperse}.B). At low $\phi$ the emulsion displays a Newtonian response, while a yield stress emerges above $\phi \simeq 0.66 \pm 0.005$, identifying the hard-sphere jamming point. This value is slightly larger than that of monodisperse frictionless spheres ($\phi_J \simeq 0.64$) and lower than for the more polydisperse silicone-oil emulsion, consistent with the reduced polydispersity.

From the yield stress $\tau_y(\phi)$ we extract the compressibility relation $\phi_J'(\Pi=\tau_y/\mu_c)$ (Fig.~\ref{SI-fig:monodisperse}.C), using $\mu_c=0.0602$ measured for the silicone-oil emulsion and imposing $\phi_J'(\Pi\to0)=\phi_J$. This yields a compressibility exponent $\alpha=0.5$, slightly larger than in the more polydisperse system. To test the robustness of the framework, we then perform the full scaling analysis allowing all parameters $(\mu_c, D, \gamma, \phi_J, A, \alpha)$ to vary. Despite these uncertainties, the collapse of the data (Fig.~\ref{SI-fig:monodisperse}.D,E) yields $\beta \simeq 2.4$, with fitted parameters close to those obtained for the silicone-oil emulsion (Figure~\ref{SI-fig:tabel}, soybean oil). The resulting constitutive laws accurately reproduce the measured $\phi$-imposed flow curves in Fig.~\ref{SI-fig:monodisperse}.B, demonstrating the robustness of the framework and the weak sensitivity of the exponent $\beta$ and $\alpha$ to droplet polydispersity.

\section*{SI.4 Osmotic pressure $\Pi$, not shear stress $\tau$, is the central variable.}

\begin{figure}[tbhp]
\centering
\includegraphics[width=14cm]{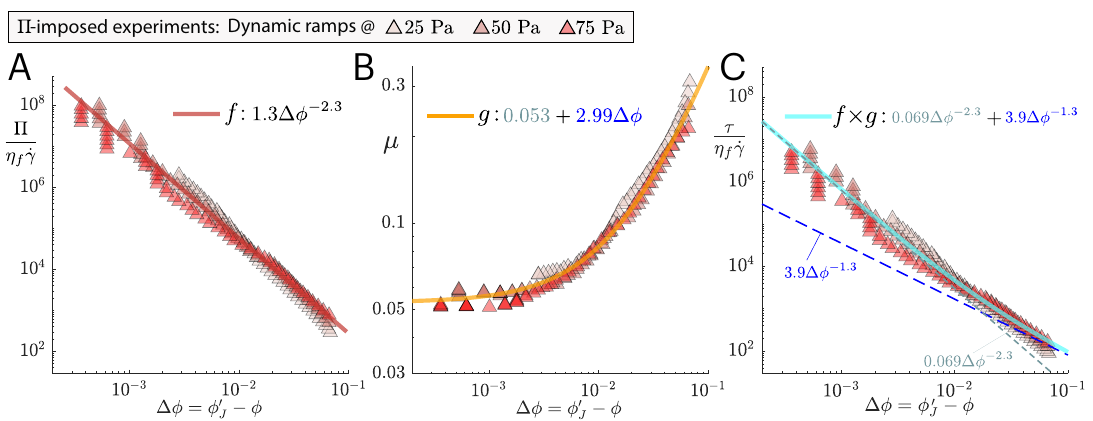}
\caption{Data are from the $\Pi$-imposed measurements in Fig.~1 of the main text.  
(A) Rescaled pressure $\Pi/\eta_f\dot{\gamma}$ versus distance to jamming $\Delta\phi = \phi_J' - \phi$, fitted by a single power law (red line). (B) Friction coefficient $\mu$ versus $\Delta\phi$, recast from Fig.~1.C of the main text, fitted by $\mu = 0.053 + 2.99\,\Delta\phi$ (orange line). 
(C) Resulting $\tau/\eta_f\dot{\gamma}=\mu \Pi/\eta_f\dot{\gamma}$ versus $\Delta\phi$, which combines two power laws.}
\label{SI-fig:collapsing1}
\end{figure}

Fig.~\ref{SI-fig:collapsing1} presents the $\Pi$-imposed measurements. The rescaled pressure $\Pi/\eta_f \dot{\gamma}$ versus distance to jamming  $\Delta \phi=\phi_J'-\phi$ collapses onto a single power law, $1.3\,\Delta\phi^{-2.3}$ (red line in panel~A). However, $\mu$ is neither constant nor a pure power law, but rather the sum of a constant and a power law: $\mu = 0.053 + 2.99\,\Delta\phi$ (orange line in panel~B). Consequently, the resulting rescaled stress $\tau/\eta_f \dot{\gamma}= \mu \times \Pi/\eta_f \dot{\gamma}$ is not a single power law but the sum of two contributions: $0.069\,\Delta\phi^{-2.3}$ and $3.9\,\Delta\phi^{-1.3}$ (cyan and blue in panel~C, respectively).

\section*{SI.5 Derivation of the $\phi$-imposed rheology and scaling analyses}

$\Pi$-imposed measurements (Fig.~1.C in the main text) show that
\begin{align}
    \phi &= \phi_J' - B^{1/\beta} J^{1/\beta} \label{SI-eqSI:phi_J}, \\
    \mu &= \mu_c + D B^{\gamma/\beta} J^{\gamma/\beta} \label{SI-eqSI:mu_J} .
\end{align}
Canceling J, we get 
\begin{equation}
    \mu(\Delta\phi) = \mu_c + D {\Delta\phi}^\gamma, \quad \text{where }  \Delta \phi = \phi_J' - \phi. \label{SI-eqSI:mu_phi}
\end{equation}
From quasi-static pressure steps (Fig.~2 in the main text), we find 

\begin{align}
    \mu_c &\approx 0.06 \label{SI-eqSI:mu_c},\\
    \phi_J' &= \phi_J + A (\frac{\Pi}{E})^\alpha. \label{SI-eqSI:phi_J_prime}
\end{align}
The set of Eq.~\ref{SI-eqSI:phi_J},~\ref{SI-eqSI:mu_J},~\ref{SI-eqSI:mu_c} and~\ref{SI-eqSI:phi_J_prime} form the constitutive laws for the rheology of emulsions. These constitutive laws allow us to predict the $\phi$-imposed rheology, and perform scaling analyses: first by expressing rheological quantities as a function of the distance to the pressure-dependent jamming volume fraction $\Delta \phi= \phi_J' - \phi$, and then by rescaling them with respect to the distance to the hard-sphere jamming limit $| \phi_J-\phi |$. 

\subsection{Prediction of $\phi$-imposed rheology}

To predict the $\phi$-imposed rheology, we solve for the shear stress $\tau$ as a function of the shear rate $\dot{\gamma}$ at fixed $\phi$.  
Starting from $\Pi = \tau / \mu(\Delta \phi)$ and substituting Eq.~\ref{SI-eqSI:mu_phi} into Eq.~\ref{SI-eqSI:phi_J_prime}, we obtain
\begin{align*}
    \phi_J' &= \phi_J + A \left( \frac{\tau}{E \mu(\Delta \phi)} \right)^\alpha \\
            &= \phi_J + A \left( \frac{\tau}{E \big(\mu_c + D (\Delta \phi)^\gamma \big)} \right)^\alpha.
\end{align*}
Rearranging gives
\begin{align}
    \Delta \phi = \phi_J - \phi + A \left( \frac{\tau}{E \big(\mu_c + D (\Delta \phi)^\gamma \big)} \right)^\alpha. 
    \label{SI-eqSI:delta_phi}
\end{align}
For a fixed $\phi$, with parameters $\{\phi_J, A, \alpha\}$ set from the compressibility law and $\{\mu_c, D, \gamma\}$ set from $\Pi $-imposed measurements, Eq.~\ref{SI-eqSI:delta_phi} can be solved numerically for $\Delta \phi$ at any $\tau$. 
Given that $J = \eta_f \dot{\gamma} / \Pi$, Eq.~\ref{SI-eqSI:phi_J} can be rewritten as
\begin{equation}
    \dot{\gamma} = \frac{\Pi (\phi_J' - \phi)^\beta}{B \eta_f}.
\end{equation}
Substituting $\Pi = \tau / \mu(\Delta \phi)$ yields
\begin{equation}
    \dot{\gamma} = \frac{\tau (\Delta \phi)^\beta}{\mu(\Delta \phi) B \eta_f}.
\end{equation}
Finally, using Eq.~\ref{SI-eqSI:mu_phi}, we obtain
\begin{equation}
    \dot{\gamma} = \frac{\tau (\Delta \phi)^\beta}{\big(\mu_c + D (\Delta \phi)^\gamma\big) B \eta_f}.
\end{equation}
Thus, for each $\phi$, given $\tau$ and the corresponding $\Delta \phi$, we compute $\dot{\gamma}$, yielding the $\tau(\dot{\gamma})$ relation at fixed $\phi$ plotted in Fig.~4.A in the main text.

\subsection{Granular-like scaling}

Rheological quantities are expressed here as a function of the distance to the pressure-dependent jamming point, $\Delta\phi = \phi_J' - \phi$. 
Since $J = \eta_f \dot{\gamma}/\Pi$, Eq.~\ref{SI-eqSI:phi_J} can be rewritten as
\begin{equation}
\Pi = \frac{B \eta_f \dot{\gamma}}{(\phi_J' - \phi)^\beta}.
\label{SI-eqSI:P_gamma_dot}
\end{equation}
Using the definition $\tau = \mu(\Delta\phi)\Pi$, this relation can equivalently be written as
\begin{equation}
\frac{\Pi}{\eta_f \dot{\gamma}}
=
\frac{\tau/\mu(\Delta\phi)}{\eta_f \dot{\gamma}}
=
\frac{B}{(\Delta\phi)^\beta}.
\label{SI-eq:gran}
\end{equation}
This framework therefore predicts that both $\Pi/(\eta_f \dot{\gamma})$ and $(\tau/\mu)/(\eta_f \dot{\gamma})$ collapse onto the same master curve when plotted as a function of $\Delta\phi$. For $\Pi$-imposed measurements, this scaling can be tested directly since all quantities ($\Pi$, $\tau$, $\phi$, $\dot{\gamma}$) are experimentally accessible. In contrast, for $\phi$-imposed measurements only $\tau$ and $\dot{\gamma}$ are directly measured. However, for a given $\phi$, the distance to jamming $\Delta\phi$ can be reconstructed numerically using Eq.~\ref{SI-eqSI:delta_phi}, once the parameters $\{\phi_J, A, \alpha\}$ (determined from the compressibility law) and $\{\mu_c, D, \gamma\}$ (determined from the $\Pi$-imposed measurements) are fixed. This procedure allows Eq.~\ref{SI-eq:gran} to be evaluated for $\phi$-imposed data as well, leading to the collapse shown in Fig.~4B.

\subsection{Two rheological branches above and below $\phi_J$}
Rheological quantities are here expressed as a function of the distance to the hard sphere jamming point $\phi_J$. Substituting Eq.~\ref{SI-eqSI:phi_J_prime} into Eq.~\ref{SI-eqSI:P_gamma_dot}, we obtain
\begin{align*}
    \Pi  &= \frac{B \eta_f \dot{\gamma}}{\left(\phi_J + A \left(\frac{\Pi }{E}\right)^\alpha - \phi\right)^\beta},  \\
    \Pi  &= \frac{B \eta_f \dot{\gamma}}{|\phi_J - \phi|^\beta \left[ \pm 1 + \frac{A}{|\phi_J - \phi|} \left(\frac{\Pi }{E}\right)^\alpha\right]^\beta}, \\
    \frac{\Pi}{E|\phi_J - \phi|^{1/\alpha}} &= \frac{B \eta_f \dot{\gamma}}{E|\phi_J - \phi|^{\beta + 1/\alpha} \left[ \pm 1 + A(\frac{\Pi }{E |\phi_J - \phi|^{1/\alpha}})^\alpha\right]^\beta}. 
\end{align*}
Then we get
\begin{align}
    Y = \frac{B X}{\left( \pm 1 + A Y^\alpha \right)^\beta},   \quad \text{where } X = \frac{\eta_f\dot{\gamma}}{E|\phi_J - \phi|^{\beta + 1/\alpha}}, Y = \frac{\Pi }{E|\phi_J - \phi|^{1/\alpha}}
\end{align}
We expect that the rescaled \(\Pi \) vs. \(\dot{\gamma}\) collapses into two branches depending on the sign of $\phi_J-\phi$, reflecting the behavior of the emulsion above and below $\phi_J$. From Eqs.~\ref{SI-eqSI:phi_J}, \ref{SI-eqSI:mu_J}, and \ref{SI-eqSI:phi_J_prime}, we derive the following asymptotic limits. For convenience, we introduce the notation $Y_\Pi = \frac{\Pi}{E|\phi_J - \phi|^{1/\alpha}}$ and $Y_\tau = \frac{\tau}{E|\phi_J - \phi|^{1/\alpha}}$.

\begin{itemize}

\item  \textbf{Newtonian regime} ($\phi < \phi_J$ and $\dot{\gamma} \to 0$):

For \(\phi < \phi_J\) in the zero shear rate limit, \(\Pi  \rightarrow 0\), so Eq.~\ref{SI-eqSI:phi_J_prime} gives \(\phi_J' = \phi_J\), Eq.~\ref{SI-eqSI:mu_phi} gives \(\mu = \mu_c + D (\phi_J - \phi)^{\gamma}\), and Eq.~\ref{SI-eqSI:phi_J} gives \(\phi = \phi_J - B^{1/\beta}(\eta_f \dot{\gamma}/\Pi )^{1/\beta}\). Altogether, we obtain
\begin{align}
    \Pi  &= \frac{B \eta_f \dot{\gamma}}{(\phi_J - \phi)^\beta} 
    &\quad \to \quad&
    Y_\Pi  = B X,  \label{SI-eqSI:normalized-P-below-jamming}\\
    \tau/\Pi  &= \mu = \mu_c + D (\phi_J - \phi)^\gamma
    &\quad \to \quad&
    Y_\tau = B X \big( \mu_c + D (\phi_J - \phi)^\gamma ). 
\end{align}

\item \textbf{Yield-stress regime} ($\phi > \phi_J$ and $\dot{\gamma} \to 0$):

For \(\phi > \phi_J\) in the zero shear rate limit, from Eq.~\ref{SI-eqSI:phi_J} and~\ref{SI-eqSI:mu_J} we have \(\phi = \phi_J', \mu = \mu_c\), and from Eq.~\ref{SI-eqSI:phi_J_prime} we obtain \(\phi = \phi_J' = \phi_J + A(\Pi _y/E)^{\alpha}\), which yields
\begin{align}
    \Pi _y/E &= \left(\frac{\phi - \phi_J}{A}\right)^{1/\alpha} 
    &\quad \to \quad&
    Y_\Pi  = A^{-1/\alpha},  \label{SI-eqSI:normalized-P-above-jamming} \\
    \tau_y/\Pi _y &= \mu = \mu_c
    &\quad \to \quad&
    Y_\tau = \mu_c A^{-1/\alpha}.   \label{SI-eqSI:yield_tau} 
\end{align}

\item \textbf{Shear-thinning regime} ($\phi \to \phi_J$ and large $\dot{\gamma}$):

Close to jamming, i.e., \(\phi \rightarrow \phi_J\), from Eq.~\ref{SI-eqSI:mu_phi} we get \(\mu = \mu_c + D (\phi_J' - \phi_J)^{\gamma} = \mu_c + D A^{\gamma}(\Pi/E)^{\alpha\gamma}\). We also obtain \(\phi_J = \phi_J' - B^{1/\beta}J^{1/\beta}\) and \(\phi_J' = \phi_J + A(\Pi/E)^{\alpha}\) from Eq.~\ref{SI-eqSI:phi_J} and Eq.~\ref{SI-eqSI:phi_J_prime}, respectively, so \(B^{1/\beta}J^{1/\beta} = A(\Pi/E)^{\alpha}\). Given $J = \eta_f \dot{\gamma}/\Pi $, we have
\begin{align}
    \Pi /E &= A^{-\frac{\beta}{\alpha\beta + 1}}(B \eta_f \dot{\gamma}/E)^{\frac{1}{\alpha\beta  +1}} 
    &\quad \to \quad&
    Y_\Pi  = A^{-\frac{\beta}{\alpha\beta + 1}} (B X)^{\frac{1}{\alpha\beta + 1}} \label{SI-eqSI:normalized-P-at-jamming}, \\
    \tau/\Pi  &= \mu = \mu_c + D A^{\gamma}(\Pi /E)^{\alpha\gamma}  
    &\quad \to \quad&
    Y_\tau = A^{-\frac{\beta}{\alpha\beta + 1}} \left( B X\right)^{\frac{1}{\alpha\beta  +1}}[\mu_c + D A^{\frac{\gamma}{\alpha\beta + 1}}(B X)^{\frac{\alpha\gamma}{\alpha\beta + 1}}|\phi_J - \phi|^{\gamma}] \label{SI-eqSI:normalized-tau-at-jamming}.
\end{align}
\end{itemize}
These equations indicate the existence of a yield stress and yield pressure when $\phi > \phi_J$. Close to jamming, the pressure follows a simple power-law relationship (Eq.~\ref{SI-eqSI:normalized-P-at-jamming}) with the shear rate, whereas the stress does not (Eq.~\ref{SI-eqSI:normalized-tau-at-jamming}). By summing the yield (Eq.~\ref{SI-eqSI:normalized-P-above-jamming}) and shear thinning (Eq.~\ref{SI-eqSI:normalized-P-at-jamming}) asymptotic limits, for $\phi > \phi_J$, the pressure can be expressed as a HB law (Eq. 7 in Main), with a shear-thinning exponent of $1/(\alpha\beta + 1)$, but not the shear stress Eq.~\ref{SI-eq:HBforTau}:
\begin{align}
\Pi/E  &= \left(\frac{\phi - \phi_J}{A}\right)^{1/\alpha} + A^{-\frac{\beta}{\alpha\beta + 1}}(B \eta_f \dot{\gamma}/E)^{\frac{1}{\alpha\beta  +1}}
&\quad \to \quad&
Y_\Pi  = A^{-1/\alpha} + A^{-\frac{\beta}{\alpha\beta + 1}} (B X)^{\frac{1}{\alpha\beta + 1}}. \label{SI-eqSI:HB-for-P} 
\end{align}
With the shear stress $\Pi_1$ in the Newtonian (Eq.~\ref{SI-eqSI:normalized-P-below-jamming}) and the shear stress $\Pi_2$ in shear-thinning (Eq.~\ref{SI-eqSI:normalized-P-at-jamming}) asymptotic limits, for $\phi < \phi_J$, we obtain a Cross-type equation: 
\begin{align}
    \Pi/E &= \frac{\Pi_1/E}{1 + \Pi_1/\Pi_2} \notag \\ 
    &= \frac{ B \eta_f \dot{\gamma}/E  }{(\phi_J - \phi)^{\beta} + A^{\frac{\beta}{\alpha\beta + 1}} (B \eta_f \dot{\gamma}/E)^{\alpha\beta/(\alpha\beta+ 1)} } 
    \label{SI-eq:Cross_type-for-P}
    &\quad \to \quad&
    Y_\Pi = \frac{BX}{1 + A^{\frac{\beta}{\alpha\beta + 1}}(BX)^{\alpha\beta/(\alpha\beta+ 1)}}
\end{align}

\begin{figure}[htbp]
\centering\includegraphics[width=1\textwidth]{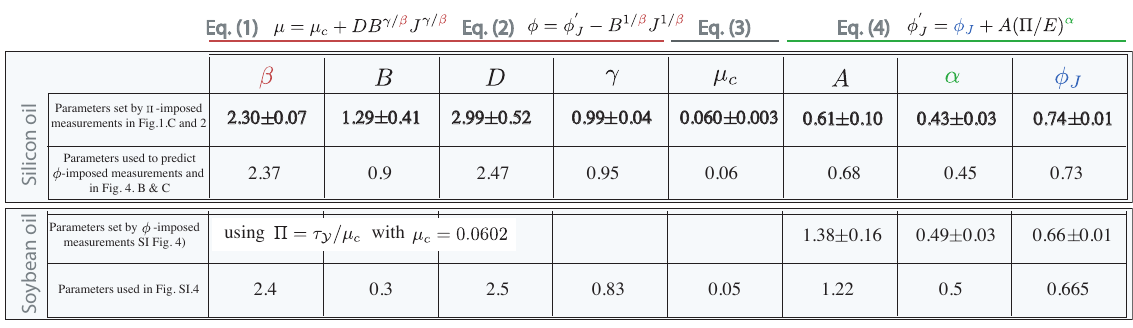}
\caption{Values of the parameters $\beta, B, \mu_c, D, \gamma, \phi_J, A, \alpha$.}
\label{SI-fig:tabel}
\end{figure}

\section*{SI.6 Parameters of the pressure-imposed framework}

For the silicone oil emulsion, the parameters $\beta, B, \mu_c, D, \gamma, \phi_J, A,$ and $\alpha$ listed in Fig.~\ref{SI-fig:tabel} (Silicon oil, top) are obtained by fitting the $\Pi$-imposed data shown in Figs.~1C and~2 of the main text. The $\phi$-imposed rheology and the collapse analyses presented in Fig.~4 are then generated by varying these parameters only \emph{within their respective uncertainty ranges}, leading to the (very close) values reported in Fig.~\ref{SI-fig:tabel} (Silicon oil, bottom). Allowing the parameters to float within their respective error bars was sufficient to obtain excellent agreement between the different experimental control parameters. The uncertainty in $\phi_J$ was $ \sim 1\%$ owing to variations in sample preparation and gap filling in the sample cell. 

For the soybean oil emulsion, $\Pi$-imposed measurements are not available. In this case, the parameters are extracted from the collapse shown in Fig.~\ref{SI-fig:monodisperse}.D and are subsequently used to predict the $\phi$-imposed rheology and to obtain the collapse in Fig.~\ref{SI-fig:monodisperse}.E.

\section*{SI.7 Osmotic pressure $\Pi$, not shear stress $\tau$, obeys Herschel-Bulkley rheology}

\begin{figure}[tbhp]
\centering
\includegraphics[width=0.9\textwidth]{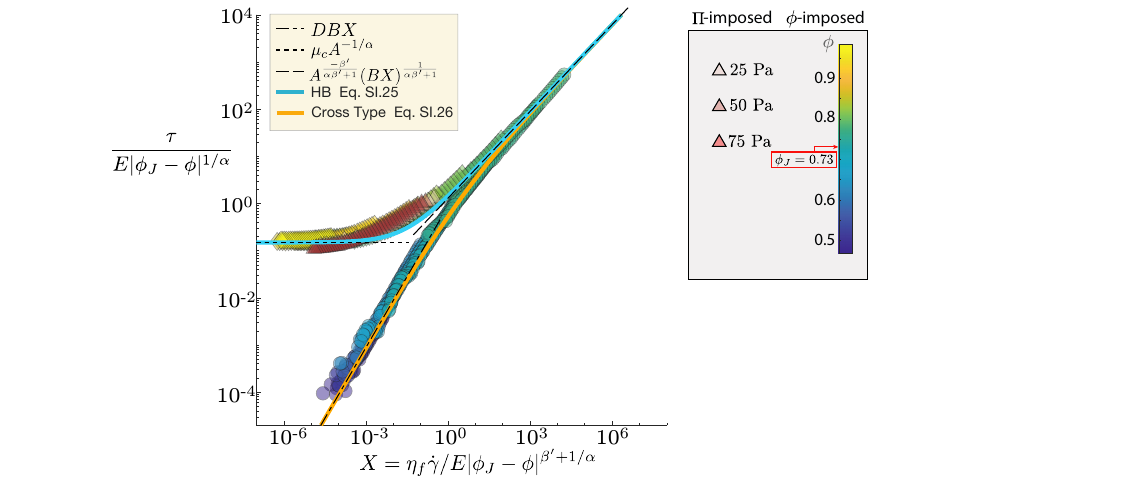}
\caption{Rescaled shear stress $\tau$ versus scaled shear rate $\dot{\gamma}$. This better collapse of $\tau$ is obtained using $\beta'=1.40$.}
\label{SI-fig:collapsing2}
\end{figure}

As shown in the inset of Fig.~4C of the main text, rescaling the shear stress
$\tau$ in the same manner as the pressure $\Pi$ does not lead to a satisfactory
data collapse. This negative result is expected within the pressure-imposed
framework. Indeed, $\tau=\mu \Pi$, where the friction coefficient $\mu$ depends on the distance to jamming, $\Delta\phi=\phi_J'(\Pi)-\phi$ (Fig.~3B), which depends on $\Pi$ and therefore on $\tau$ itself. Combining the asymptotic expressions derived in
Eqs.~\ref{SI-eqSI:yield_tau} and~\ref{SI-eqSI:normalized-tau-at-jamming}, we obtain
\begin{align}
\tau/E &= \mu_c \left(\frac{\phi - \phi_J}{A}\right)^{1/\alpha}
+ A^{-\frac{\beta}{\alpha\beta + 1}}
\left(B \eta_f \dot{\gamma}/E\right)^{\frac{1}{\alpha\beta +1}}
\left[\mu_c + D A^{\frac{\gamma}{\alpha\beta + 1}}
\left(B\eta_f \dot{\gamma}/E\right)^{\frac{\alpha\gamma}{\alpha\beta + 1}}\right],
\label{SI-eq:HBforTau}
\end{align}
which explicitly demonstrates that $\tau$ cannot be written as the sum of a
constant yield stress and a simple power law of the shear rate. Consequently,
the shear stress does not follow a Herschel--Bulkley form, in agreement with
previous findings for soft-sphere systems~\cite{kawasaki2015diverging,tapia2024rheology}.

As shown in Fig.~\ref{SI-fig:collapsing2}, an apparent collapse of $\tau$ can nonetheless be achieved by treating the exponent $\beta$ as a free fitting parameter (with $\alpha$ and $\phi_J$ fixed by the compressibility law). The collapse yields $\beta^{\prime}=1.4$, a value inconsistent~--~beyond experimental uncertainty~--~with the value $\beta=2.30 \pm 0.07$ set by the pressure imposed flow rules. Such a collapse of  $\tau$ implies the three asymptotic regimes must each be approximated by either a constant or a power law, which rationalizes the value of $\beta^{\prime}=1.4$. In the Newtonian regime, since $\mu_c$ is negligible, one obtains $\mu \approx D\Delta\phi^\gamma \approx D(\phi_J' - \phi)^\gamma$, yielding 
\begin{align}
    \tau_{|_{\rm{Newt}}} &= \mu \Pi \approx \frac{D B \eta_f \dot{\gamma}}{(\phi_J' - \phi)^{\beta - \gamma}}
    \approx \frac{D B \eta_f \dot{\gamma}}{(\phi_J' - \phi)^{\beta^{\prime}}}, 
    \label{SI-eqSI:tau-below-jamming-approx}
\end{align}
with an effective exponent $\beta^{\prime} \approx \beta - \gamma = 1.4$.  In the yielding regime, $\mu = \mu_c$, leading to the yield stress given by Eq.~\ref{SI-eqSI:yield_tau}. In the shear-thinning regime, the shear stress (Eq.~\ref{SI-eqSI:normalized-tau-at-jamming}) is the sum of two power laws with similar exponents, which over the experimentally accessible range can be well approximated by a single effective power law, analogous to Eq.~\ref{SI-eqSI:normalized-tau-at-jamming} but with $\beta$ replaced by $\beta^{\prime}$: 
\begin{equation}
\frac{\tau}{E}_{|_{\rm{ST}}} \approx A^{-\frac{\beta^{\prime}}{\alpha\beta^{\prime} + 1}}\left(B \eta_f \dot{\gamma}/E\right)^{\frac{1}{\alpha\beta^{\prime} +1}}. 
\label{SI-eqSI:tau-at-jamming-approx}
\end{equation}
Similar stress-based collapses have been proposed in earlier studies of emulsions and soft spheres~\cite{hatano2008scaling,PhysRevE.80.011308,nordstrom2010microfluidic,tighe2010model,paredes2013rheology,dinkgreve2015universal,bonn2017yield,velandia2025rescaling}, but they should be regarded as approximate~\cite{kawasaki2015diverging,tapia2024rheology}. For $\phi > \phi_J$, combining (Eq.~\ref{SI-eqSI:yield_tau}) and (Eq.~\ref{SI-eqSI:tau-at-jamming-approx}), we obtain an approximate Herschel-Bulkley law for the shear stress (blue line in Fig. \ref{SI-fig:collapsing2}), 
\begin{equation}
\frac{\tau}{E}_{|_{\phi>\phi_J}} \approx \mu_c \left(\frac{\phi - \phi_J}{A}\right)^{1/\alpha} + A^{-\frac{\beta^{\prime}}{\alpha\beta^{\prime} + 1}}\left(B \eta_f \dot{\gamma}/E\right)^{\frac{1}{\alpha\beta^{\prime} +1}}.
\label{SI-eqSI:HB-for-tau-approx}
\end{equation}
For $\phi < \phi_J$, using the shear stress $\tau_1$ in the Newtonian (Eq.~\ref{SI-eqSI:tau-below-jamming-approx}) and $\tau_2$ in shear-thinning (Eq.~\ref{SI-eqSI:tau-at-jamming-approx}) asymptotic limits, we obtain the following approximate \emph{Cross-type} equation (orange line in Fig. \ref{SI-fig:collapsing2}): 
\begin{equation}
\frac{\tau}{E}_{|_{\phi<\phi_J}} \approx \frac{\tau_1/E}{1 + \tau_1/\tau_2} \approx \frac{ D B \eta_f \dot{\gamma}/E  }{(\phi_J - \phi)^{\beta^{\prime}} + D A^{\frac{\beta^{\prime}}{\alpha\beta^{\prime} + 1}} (B \eta_f \dot{\gamma}/E)^{\alpha\beta^{\prime}/(\alpha\beta^{\prime} + 1)}}.
    \label{SI-eq:Cross_type-for-tau-approx}
\end{equation}

By contrast, the collapse of the pressure $\Pi$ follows directly and exactly from the constitutive laws, highlighting pressure as the natural organizing variable of the flow. As explained in the main text (Eq. {7}), for $\phi>\phi_J$, summing the yield and shear-thinning asymptotic contributions shows that the osmotic pressure $\Pi$ obeys the Herschel-Bulkley (HB) law (blue line in Fig.~4.C)
\begin{equation}
\frac{\Pi}{E}_{|_{\phi>\phi_J}} =  \left(\frac{\phi - \phi_J}{A}\right)^{1/\alpha}+A^{-\frac{\beta}{\alpha\beta + 1}}(B \eta_f \dot{\gamma}/E)^{\frac{1}{\alpha\beta  +1}}.
    \label{SI-eq:HBforP}
\end{equation}
For $\phi<\phi_J$, with the pressure $\Pi_1$ in the Newtonian (Eq.~\ref{SI-eqSI:normalized-P-below-jamming}) and the pressure $\Pi_2$ in shear-thinning (Eq.~\ref{SI-eqSI:normalized-P-at-jamming}) asymptotic limits, we obtain the \emph{Cross-type} equation (orange line in Fig.~4.C) 
\begin{equation}
\frac{\Pi}{E}_{|_{\phi<\phi_J}} = \frac{\Pi_1/E}{1 + \Pi_1/\Pi_2} = \frac{ B \eta_f \dot{\gamma}/E  }{(\phi_J - \phi)^{\beta} + A^{\frac{\beta}{\alpha\beta + 1}} (B \eta_f \dot{\gamma}/E)^{\alpha\beta/(\alpha\beta + 1)} } 
    \label{SI-eq:Cross_type}
\end{equation}
Eqs.~\ref{SI-eqSI:HB-for-tau-approx}, \ref{SI-eq:Cross_type-for-tau-approx}, \ref{SI-eq:HBforP}, and \ref{SI-eq:Cross_type} provide useful explicit predictions for the shear stress and the pressure applicable under $\phi$-imposed conditions and covering the entire range of regimes, from Newtonian, yield and shear thinning.

\end{document}